\documentclass[a4paper,accepted=2025-11-17]{quantumarticle}
\pdfoutput=1
\usepackage[utf8]{inputenc}
\usepackage[english]{babel}
\usepackage[T1]{fontenc}

\usepackage{graphicx}
\usepackage{hyperref}
\usepackage[linesnumbered, ruled, vlined, titlenotnumbered, noend]{algorithm2e}
\usepackage{amsmath, amsfonts, amssymb, amsthm, xcolor}
\usepackage{tikz-cd}
\usepackage[capitalise]{cleveref}
\crefname{algocf}{alg.}{algs.}
\Crefname{algocf}{Algorithm}{Algorithms}

\newcommand{\ket}[1]{|#1\rangle}

\newcommand{\R}{{\mathbb{R}}}

\newcommand{\HH}{{\bf{H}}}
\newcommand{\ndataqubits}[0]{n}
\newcommand{\nancillas}[0]{n_a}

\newcommand{\nrounds}[0]{r}
\newcommand{\plog}[0]{p_{\log}}
\newcommand{\chainmodel}[3]{{\bf Chain}(#1, #2, #3)}

\usepackage[numbers,sort&compress]{natbib}

\begin{document}

\title{Quantum error correction for long chains of trapped ions}
\author{Min Ye}
\author{Nicolas Delfosse}
\affiliation{IonQ Inc.}

\begin{abstract}
We propose a model for quantum computing with long chains of trapped ions and we design quantum error correction schemes for this model.
The main components of a quantum error correction scheme are the quantum code and a quantum circuit called the syndrome extraction circuit, which is executed to perform error correction with this code.
In this work, we design syndrome extraction circuits tailored to our ion chain model, a syndrome extraction tuning protocol to optimize these circuits, and we construct new quantum codes that outperform the state-of-the-art for chains of about $50$ qubits.
To establish a baseline under the ion chain model, we simulate the performance of surface codes and bivariate bicycle (BB) codes equipped with our optimized syndrome extraction circuits.
Then, we propose a new variant of BB codes defined by weight-five measurements, that we refer to as BB5 codes and we identify BB5 codes that achieve a better minimum distance than any BB codes with the same number of logical qubits and data qubits, such as a $[[48, 4, 7]]$ BB5 code.
For a physical error rate of $10^{-3}$, the $[[48, 4, 7]]$ BB5 code achieves a logical error rate per logical qubit of $5 \cdot 10^{-5}$, which is four times smaller than the best BB code in our baseline family.
It also achieves the same logical error rate per logical qubit as the distance-7 surface code but using four times fewer physical qubits per logical qubit.
\end{abstract}

\maketitle

\section{Introduction}

Fault-tolerant quantum computing relies on extensive quantum error correction to correct faults occurring during the computation before they spread to all the qubits.
To maximally benefit from quantum error correction, it is crucial to design quantum error correction schemes that are tailored to the specific constraints of the hardware.
Surface codes~\cite{dennis2002topological, fowler2012surface} and color codes~\cite{bombin2006topological}, which are defined by local measurements on a lattice of qubits, naturally fit with superconducting chips, consisting of a square grid of qubits with nearest neighbor connectivity~\cite{ chen2022calibrated, krinner2022realizing, zhao2022realization, google2023suppressing, ai2024quantum, lacroix2024scaling}.
Quantum error correction schemes have been optimized for several other hardware platforms such as Majorana qubits~\cite{chao2020optimization, paetznick2023performance}, neutral atoms~\cite{bluvstein2024logical, xu2024constant, veroni2024optimized, reichardt2024logical, pecorari2025high}, cat qubits~\cite{guillaud2019repetition, ruiz2025ldpc, putterman2025hardware}, photonic qubits~\cite{bartolucci2023fusion, de2024spin, hilaire2024enhanced, walshe2025linear} or short ion chains with shuttling~\cite{ryan2021realization, ryan2024high, paetznick2024demonstration}.

In this work, we investigate the question of designing quantum error correction schemes for a long chain of trapped ions, which is the core module of a trapped ion quantum computer.
The potential of long ion chains has been demonstrated experimentally over the past 10 years with 5-ion chains running Shor's algorithm~\cite{monz2016realization} or Deutsch–Jozsa and Bernstein-Vazirani algorithms~\cite{debnath2016demonstration},
a 14-ion GHZ state~\cite{monz201114}, a quantum chemistry algorithm executed on a 15-ion chain~\cite{nam2020ground},
and quantum error correction experiments with chains of 13 to 16 ions~\cite{egan2021fault, egan2021scaling, postler2022demonstration}.
The performance of a 30-qubit quantum computer based on a long chain of ions is analyzed in~\cite{chen2024benchmarking}.
Chains with more than 50 ions have been used for analog quantum simulations~\cite{zhang2017observation, kranzl2022controlling}.
The loading of a chain of 155 ions is reported in~\cite{kamsap2017experimental}.
Architectures for large-scale quantum computers based on ion chains connected through shuttling or optical interconnects are proposed in~\cite{kielpinski2002architecture, monroe2014large, schwerdt2024scalable}.
For a review of trapped ion quantum computing see~\cite{bruzewicz2019trapped}.

The major advantages of long ion chains is their large coherence time and high connectivity.
Current trapped ion technologies achieve a gate error rate close to $10^{-3}$ for two-qubit gates and $10^{-4}$ for single-qubit gates~\cite{egan2021fault, chen2024benchmarking, decross2024computational, loschnauer2024scalable}.
Moreover, qubits stored in a long ion chain are fully connected, in the sense that the native gate set includes entangling gates for every pair of qubits~\cite{sorensen1999quantum, sorensen2000entanglement}.
This high connectivity suggests that trapped ions could be a good fit for the implementation of quantum low-density parity-check (LDPC) codes.
Recent simulations show that quantum LDPC codes achieve a significantly lower overhead than surface codes while maintaining the same logical error rate~\cite{tremblay2022constant, bravyi2024high, berthusen2025toward}.
However, these codes require some form of long-range connectivity that is not available in today's superconducting chips.
This makes long ion chains a natural candidate to host quantum LDPC codes, but a detailed analysis is needed to confirm this intuition.
There are major differences between the model considered in previous simulations of quantum LDPC codes and the capability of trapped ion qubits. These differences could translate to a large gap between simulation results and a real implementation of these codes.
For example, the fitting parameters for surface code's logical error rate derived from existing circuit models differ significantly from those obtained using our new model, with the latter more accurately reflecting ion-trap architecture performance. See \cref{sec:fitting_formulas} for a detailed discussion.

To close this gap, we propose the {\em ion chain model}, which captures the main features of quantum computing with a long ion chain. 
Namely, 
(i) idle qubits have very long coherence time,
(ii) two-qubit operations are noisier than single-qubit operations, which are noisier than idle qubits,
(iii) qubits are fully connected,
(iv) unitary gates are sequential: One can only implement a single unitary gate at a time,
(v) reset and measurement are parallel: One can reset or measure any subset of qubits in parallel,
(vi) measurements are slower than other operations.
Property (i) and (iii) suggest that quantum LDPC codes should be a good fit for such a quantum platform. 
However, one could be concerned that (iv) and (vi) may result in a significant degradation of the code performance, making quantum LDPC codes unsuitable for this model, and these limitations are not reflected in the quantum LDPC code simulations cited above.

We study the performance of quantum error correction codes encoding $k$ logical qubits into $n$ data qubits with $n \leq 50$.
Including ancilla qubits, all the quantum error correction schemes designed in this paper use a chain of at most 57 qubits.
The code parameters are denoted $[[n, k]]$, or $[[n, k, d]]$ if the minimum distance $d$ is known, and the number of ancilla qubits is denoted $n_a$.
The minimum distance $d$ measures the error correction capability of the code. 

Quantum error correction is implemented by executing a quantum circuit called the syndrome extraction circuit. 
We design syndrome extraction circuits respecting the constraints of the ion chain model (\cref{algorithm:ion_chain_syndrome_extraction}), that can be used for any stabilizer codes~\cite{gottesman1997stabilizer}.
The high connectivity of the ion chain model offers a lot of flexibility in the construction of this circuit.
To exploit this freedom, we propose a syndrome extraction tuning protocol (\cref{algorithm:syndrome_extraction_tunning_protocol}) that optimizes the syndrome extraction circuit to achieve a favorable tradeoff between logical error rate and qubit overhead.

To establish a baseline for the ion chain model, we consider surface codes, which achieve state-of-the-art performance for superconducting architecture~\cite{dennis2002topological, fowler2012surface}, and bivariate bicycle (BB) codes, which achieve state-of-art-performance when long-range gates are available~\cite{bravyi2024high}.
We simulate the performance of these codes with a syndrome extraction circuit optimized for the ion chain model.
Our baseline family includes surface codes with distance $3, 5$ and $7$.
The number of BB codes with up to 50 data qubits is massive and some of them have poor parameters. Therefore, we do not simulate all of them. Instead, we simulate a BB code with optimal minimum distance $d$ for each achievable pair of $[[n, k]]$ within the BB code family for $n \leq 50$.

Finally, we study a variant of BB codes defined by weight-5 stabilizer generators, in contrast to the weight-6 stabilizer generators of the original BB codes. 
We refer to these new codes as BB5 codes.
To avoid any confusion, in the rest of this paper we use the term BB6 codes for the original BB codes of~\cite{bravyi2024high}.
We identify two BB5 codes with parameters $[[30, 4, 5]]$ and $[[48, 4, 7]]$ that achieve larger code distance than any BB6 code with the same parameters $[[n, k]]$.
For comparison BB6 codes only achieve $[[30, 4, 4]]$ and $[[48, 4, 6]]$.
The BB5 codes we consider belong to the broader families of tensor-product generalized bicycle codes~\cite{kovalev2013quantum} and  multivariate bicycle codes proposed in~\cite{voss2024multivariate}.
In~\cite{voss2024multivariate} the authors previously discovered a code with the same parameters as our $[[30, 4, 5]]$ BB5 code by searching over a set of codes that is a subset of BB5 codes.

Our simulations show that, under the ion chain model, the $[[48, 4, 7]]$ BB5 code achieves a logical error rate per logical qubit $4$ times smaller than the best BB6 codes with code length up to $50$.
Moreover, it reaches the same logical error rate per logical qubit as the distance-7 surface code while using $4$ times fewer physical qubits per logical qubit.

The rest of this paper is organized as follows.
\cref{sec:ion_chain_model} introduces the ion chain model. 
Syndrome extraction circuits for the ion chain model are designed and optimized in \cref{sec:syndrome_extraction_circuit}.
In \cref{sec:ion_chain_codes}, we investigate the performance of surface codes and BB codes under the ion chain model and we design new codes that outperform these codes.

\section{The ion chain model}
\label{sec:ion_chain_model}

The ion chain model describes the connectivity, the parallelism and the noise rate of quantum operations in a chain of trapped ions. 
It is a useful model for machines such as the IonQ Forte system and the ion chain experiments cited in introduction.
The ion chain model, denoted $\chainmodel{n}{p}{\tau_m}$, has three parameters: $n$ for the number of qubits, $p$ for the noise parameter and $\tau_m$ which controls the measurement time.
We refer to a register of qubits for this model as a {\em $n$-qubit chain} and $p$ is called the {\em physical error rate}.
This section describes the properties of $n$-qubit chains.

A $n$-qubit chain is a register of $n$ qubits equipped with the following operations:
(i) prepare or reset any subset of qubits in the state $\ket 0$,
(ii) apply a single-qubit unitary gate to any qubit,
(iii) apply a two-qubit unitary gate to any pair of qubits,
(iv) measure any subset of qubits in the computational basis.
We assume that these operations are applied sequentially, {\em i.e.} only one of these operations can be executed at a given time step.

We adopt the standard circuit level noise model in which every operation is followed by depolarizing noise. 
A two-qubit unitary gate is followed by a random two-qubit Pauli error with probability $p$.
This error is selected uniformly among the 15 Pauli errors acting non-trivially on the support of the gate.
A single-qubit operation (preparation, reset or unitary gate) is followed by a random single-qubit Pauli error with probability $p/10$.
This error is selected uniformly among $X, Y,$ and $Z$ errors.
The outcome of a measurement is flipped with probability $p/10$.
During a gate, any idle qubit suffers from a random single-qubit Pauli error with probability $p/100$.

Empirical data supports our noise model: \cite{postler2024demonstration} reported a two-qubit gate error rate of $0.027$ and a single-qubit gate error rate of $0.0036$. Another paper \cite{chen2024benchmarking} documented single-qubit gate error rates of $2\times 10^{-4}$ and two-qubit gate error rates of $46.4\times 10^{-4}$. In both cases, single-qubit gate errors were approximately an order of magnitude smaller than two-qubit gate errors. This validates our assumption that single-qubit operations incur an error rate ten times lower than two-qubit operations.
It was further reported in \cite{postler2024demonstration} that the measurement and initialization error rates are both $0.003$, which is very close to the single-qubit gate error rate $0.0036$. This supports our choice of assigning the same error rate to single-qubit operations, measurements and initializations in our noise model.

Finally, we assume that all the operations have the same duration, except measurements which take $\tau_m$ times longer.
This long measurement time translates to more idle noise during a measurement.
We consider a regime with $\tau_m p << 100$ and we set the idling noise rate during a measurement to $\tau_m p / 100$.
In our simulations, we use $\tau_m = 30$.

In practice, the ions are confined by an electromagnetic field.
Each qubit is stored within two energy levels of an ion. 
A $n$-qubit chain may contain more than $n$ ions if we use additional ions for sympathetic cooling~\cite{myatt1997production}.
Remarkably, operations entangling any of the $\binom{n}{2}$ pairs of qubits are available as native gates in the ion chain model.
These operations can be implemented using the Mølmer–Sørensen scheme by illuminating the targeted ions using two laser beams~\cite{sorensen1999quantum, sorensen2000entanglement}.
This allows us to operate on a pair of distant qubits by controlling the laser beams without moving the ions. 
The operation is mediated by the common vibrations of the ions.
See Fig.~1 of \cite{chen2024benchmarking} for an illustration and for more details on the experimental setup.

It is worth noting that most aspects of the ion chain model have been individually highlighted in the literature. For example, the significantly slower measurement times in ion-trap systems is well-known \cite{haffner2008quantum,chamberland2018new}, sequential unitary operations have been emphasized \cite{postler2022demonstration}, and differential noise rates for two-qubit versus single-qubit operations have been documented \cite{postler2022demonstration,pogorelov2025experimental}.
However, our proposed model exhibits a critical modeling advantage over existing works. While prior studies provide only single, fixed data points derived from specific experimental conditions, our model proposes a parameterized framework based on a single variable. This allows us to generate a performance curve across a range of error rates. 
This methodology directly mirrors the widely accepted parallel circuit model (which assumes uniform noise rates for all operations) used for evaluating code performance on superconducting chips \cite{fowler2012surface,tremblay2022constant, bravyi2024high, berthusen2025toward}.
Ultimately, our model provides a comprehensive and computationally efficient tool for simulating fault-tolerant performance on ion chain hardware.

In principle, one could construct a multi-variable noise model for a more precise characterization of ion chain hardware. For example, by assigning three independent variables to denote the error rates for single-qubit operations, measurements, and initializations, instead of using a common $p/10$ rate. However, adopting a multi-variable approach would significantly increase the model's complexity, making it less convenient for the primary goal of code simulation and performance assessment.
We believe our model, parameterized by a single variable $p$, achieves a good trade-off between practical convenience and necessary accuracy.
This philosophy is exactly analogous to the widely accepted parallel circuit model used for evaluating code performance on superconducting chips. While the error rates for different operations on superconducting hardware are never precisely uniform, that model's assumption of uniform noise rates is still considered a reasonable approximation and is used extensively \cite{fowler2012surface,tremblay2022constant, bravyi2024high, berthusen2025toward}. Our model adopts the same pragmatic approach for the ion chain hardware, prioritizing comprehensive, efficient simulation over the marginal precision gained by tracking multiple independent parameters.

\section{Syndrome extraction circuits for the ion chain model}
\label{sec:syndrome_extraction_circuit}

We focus on {\em stabilizer codes} which are defined as the common $+1$-eigenspace of a family of commuting Pauli operators called {\em stabilizer generators}~\cite{gottesman1997stabilizer}.

The {\em syndrome extraction circuit} is the quantum circuit that performs the measurement of the stabilizer generators of the code.
It is executed periodically to correct errors.
The ion chain model offers a lot of flexibility when designing syndrome extraction circuits, which can be used to reduce the qubit overhead of quantum error correction.
For instance, the all-to-all qubit connectivity enables the use of a single ancilla qubit to sequentially measure all stabilizer generators for any code.
This reduces the total physical qubit count of the surface and BB6 codes by a factor of almost $2$. Note that this is not possible for quantum hardware with local connectivity constraints, where a single ancilla qubit can only measure stabilizers supported on its neighbors.
On the other hand, using a single ancilla qubit may result in a long waiting time and a suboptimal logical error rate. 
To speed up the syndrome extraction and to reduce the logical error rate, one may use multiple ancilla qubits that are measured simultaneously.
In this section, we present a family of syndrome extraction circuits with a variable number of ancilla qubits.
Note that this syndrome extraction circuit is a standard application of the Hadamard test; see for example Figure 10.13 on page~473 of \cite{nielsen2010quantum}.
Then, we propose a protocol to select the number of ancilla qubits that optimizes the tradeoff between the qubit overhead and the logical error rate.

\begin{algorithm}
\DontPrintSemicolon
\SetAlgoLined
\KwIn{
A list of $\ndataqubits$-qubit Pauli operators $S_0, S_1 \dots, S_{r-1}$ supported on qubits $0,1,\dots, \ndataqubits-1$.
An integer $\nancillas \geq 1$.
}
\KwOut{A quantum circuit measuring the operators $S_0, S_1 \dots, S_{r-1}$.}
\For{$i=0,1,\dots, \left\lceil \frac{r}{\nancillas} \right\rceil - 1$}{
    \For{$j=0,1,\dots, \nancillas-1$}{
        $k \leftarrow i \nancillas + j$.\;
        $c \leftarrow \ndataqubits + j$.\;
        Prepare qubit $c$ in the state $\ket 0$.\;
        Apply a $H$ gate to qubit $c$.\;
        \For{$t=0,1,\dots, \ndataqubits - 1$}{
            \If{the component $P$ of $S_k$ on qubit $t$ is $P=X, Y$ or $Z$}{
                Apply a controlled-$P$ gate controlled on qubit $c$ with target qubit $t$.\;            
            }
        }
        Apply a $H$ gate to qubit $c$.\;
        {\bf if} {\em $k \geq r-1$} {\bf then} break;\;
    }
    Measure qubits $n,n+1, \dots, n+j$.
}
\caption{Ion chain syndrome extraction circuit}
\label{algorithm:ion_chain_syndrome_extraction}
\end{algorithm}

The {\em ion chain syndrome extraction circuit} is described in \cref{algorithm:ion_chain_syndrome_extraction}.
This circuit fits the constraints of the ion chain model.
It measures Pauli operators supported on $\ndataqubits$ data qubits indexed by $0, 1, \dots, \ndataqubits-1$ using $\nancillas$ ancilla qubits indexed by $\ndataqubits, \ndataqubits+1, \dots, \ndataqubits + \nancillas-1$. The ancilla qubits are measured simultaneously.
The parallelization of these measurements yields a significant reduction of the total idle time of the qubits because measurements are the slowest operations in the ion chain model.
Other gates remain sequential to respect the ion chain model.

To obtain the syndrome extraction circuit of a stabilizer code with $\nrounds$ rounds of syndrome extraction, we set the input of \cref{algorithm:ion_chain_syndrome_extraction} as the stabilizer generators of the code repeated $\nrounds$ times.
In numerical simulations, we use $\nrounds = d$ rounds of syndrome extraction where $d$ is the minimum distance of the code, which is a standard choice in the literature~\cite{bravyi2024high}.
\cref{algorithm:ion_chain_syndrome_extraction} is used with an input sequence alternating $X$ and $Z$ operators.

To optimize the syndrome extraction circuit, we need to identify a sweet spot between the number of ancilla qubits consumed and the logical error rate achieved.
We propose the {\em syndrome extraction tuning protocol} described in \cref{algorithm:syndrome_extraction_tunning_protocol}.
The protocol keeps increasing the number $n_a$ of ancilla qubits as long as each iteration reduces the logical error rate by a factor $\gamma$.
A smaller value of $\gamma$ makes the while loop's condition harder to satisfy, causing the loop to exit earlier and thereby yielding a smaller chosen value for $n_a$. In \cref{tab:baseline_code_parameters}, we set $\gamma=0.9$ in \cref{algorithm:syndrome_extraction_tunning_protocol} to obtain the optimized value of $n_a$. We also ran \cref{algorithm:syndrome_extraction_tunning_protocol} for several $\gamma$ values between $0.8$ and $1$, finding that the algorithm's output is largely unaffected by $\gamma$. Specifically, for $0.8\le\gamma\le 1$, the output varies by at most $2$, and often remains consistent across different $\gamma$ values.
The optimized number of ancilla qubits $n_a$ returned by \cref{algorithm:syndrome_extraction_tunning_protocol} depends on the code, the measurement time $\tau_m$, and to a lesser extent on $p$. 
The effectiveness of \cref{algorithm:syndrome_extraction_tunning_protocol} is validated in \cref{sec:validate_tuning}.

As mentioned before, two-qubit gates in ion chains can be implemented via the Mølmer–Sørensen mechanism, where interactions are mediated by the ions' motional modes \cite{sorensen1999quantum, sorensen2000entanglement}. Generally, the Center-of-Mass (COM) mode is utilized. However, as the length of the ion chain increases, the motional modes become denser in frequency, leading to mode crowding and residual coupling to other modes \cite{haffner2008quantum, wu2018noise, monroe2021programmable}.
This residual coupling is a critical limitation because it directly increases the error rate of two-qubit gates, thereby constraining the practical length of ion chains. We note that the precise impact of this effect is beyond the scope of this paper and is not included in our simulation. If accounted for, this effect would likely further reduce the optimal number of ancilla qubits, as it increases physical error rates for longer chains.

\begin{algorithm}
\DontPrintSemicolon
\SetAlgoLined
\KwIn{
A stabilizer code $C$ with $\ndataqubits$ data qubits.
Parameters $p \in [0,1]$, $\tau_m \in \R_+$ and $\gamma \in [0,1]$.
}
\KwOut{
A positive integer $n_a$.
}
Initialize $n_a = 1$ and $\plog(0) = 1$.\;
Estimate the logical error rate $\plog(n_a)$ of $C$ with the ion chain syndrome extraction circuit using $n_a$ ancilla qubits for the model $\chainmodel{\ndataqubits+n_a}{p}{\tau_m}$.\;
\While{$\frac{\plog(n_a)}{\plog(n_a-1)} < \gamma$}{
    $n_a \leftarrow n_a + 1$.\;
    Estimate the logical error rate $\plog(n_a)$ of $C$ with the ion chain syndrome extraction circuit using $n_a$ ancilla qubits for the model $\chainmodel{\ndataqubits+n_a}{p}{\tau_m}$.\;
}
\Return{$n_a$.}
\caption{Syndrome extraction tuning protocol}
\label{algorithm:syndrome_extraction_tunning_protocol}
\end{algorithm}

\section{Improved codes for the ion chain model}
\label{sec:ion_chain_codes}

\begin{figure}
\includegraphics[width=\linewidth]{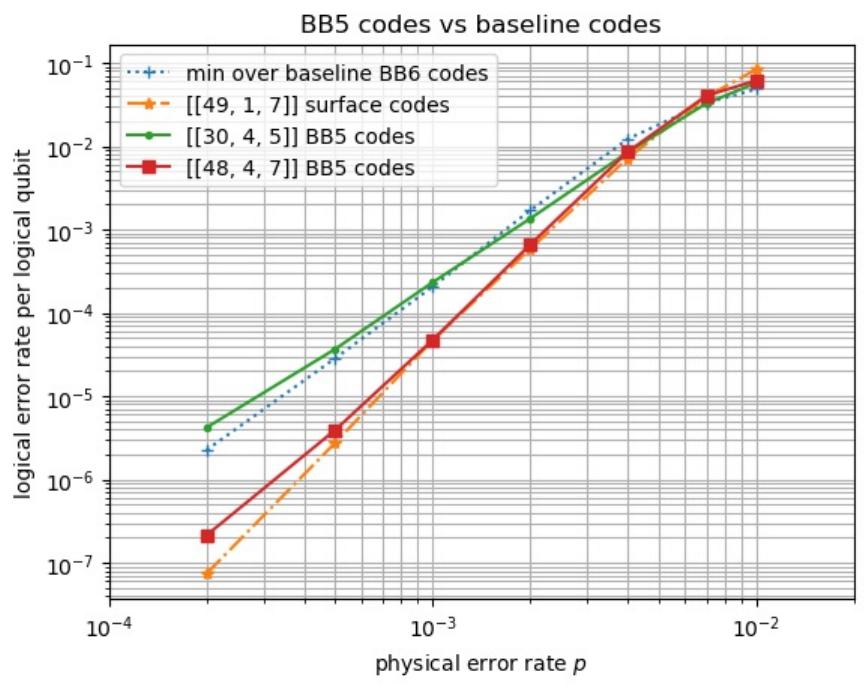}
    \caption{Comparison between BB5 codes and baseline codes. Each point on the blue curve is the minimum logical error rate per logical qubit for a BB6 code from the baseline family. More specifically, there are 11 BB6 codes in the baseline family, as listed in \cref{tab:baseline_code_parameters}. For each value of physical error rate $p$, we obtain the logical error rate per logical qubit for these 11 BB6 codes through Monte Carlo simulations, and each point on the blue curve is simply the minimum of these 11 values.}
    \label{fig:BB5_vs_baseline}
\end{figure}

As a baseline, we consider surface codes~\cite{dennis2002topological, fowler2012surface}, and BB6 codes~\cite{bravyi2024high}.
These code constructions are reviewed in \cref{appendix:surface_code_and_bb_codes}.
For our baseline family, we select surface codes with minimum distance $3, 5$ and $7$ and we select a BB6 code achieving the largest minimum distance $d$ for each achievable pair $[[n,k]]$ within BB6 code family for $n \leq 50$.
We eliminate codes with $d \leq 2$.
The parameters obtained are listed in \cref{tab:baseline_code_parameters}.

\begin{table*}
    \centering

    \begin{tabular}{|c|c|c|c|}
        \hline
        code & parameters & stabilizer weight & optimized $n_a$ \\
        \hline
        \hline
         Surface code & $[[9, 1, 3]]$ & $\leq 4$ & 4 \\
         Surface code & $[[25, 1, 5]]$ & $\leq 4$ & 5 \\
         Surface code & $[[49, 1, 7]]$ & $\leq 4$ & 8 \\
        \hline
         BB6 code & $[[18, 4, 4]]$ & 6 & 4 \\
         BB6 code & $[[24, 4, 4]]$ & 6 & 3 \\
         BB6 code & $[[28, 6, 4]]$ & 6 & 5 \\
         BB6 code & $[[30, 4, 4]]$ & 6 & 3 \\
         BB6 code & $[[30, 8, 4]]$ & 6 & 5 \\
         BB6 code & $[[36, 4, 4]]$ & 6 & 4 \\
         BB6 code & $[[36, 8, 4]]$ & 6 & 4 \\
         BB6 code & $[[42, 4, 6]]$ & 6 & 5 \\
         BB6 code & $[[42, 6, 6]]$ & 6 & 7 \\
         BB6 code & $[[48, 4, 6]]$ & 6 & 6 \\
         BB6 code & $[[48, 8, 4]]$ & 6 & 4 \\
        \hline
         BB5 code & $[[30, 4, 5]]$ & 5 & 5 \\
         BB5 code & $[[48, 4, 7]]$ & 5 & 6 \\
        \hline
    \end{tabular}
    \caption{Parameters of baseline codes and BB5 codes. The optimized $n_a$ is the number of ancilla qubits used for syndrome extraction, selected by \cref{algorithm:syndrome_extraction_tunning_protocol} with parameters $p=5\times 10^{-4}$, $\tau_m=30$ and $\gamma=0.9$.}
    \label{tab:baseline_code_parameters}
\end{table*}

Next we propose a new code family that we call BB5 codes and we show that they outperform all the baseline codes under the ion chain model.
Let $Q_\ell$ be the $\ell \times \ell$ circulant matrix whose first row is $(0 1 0 \dots 0)$.
A BB5 code is a CSS code defined by the pair of matrices
\begin{align*}
&\HH_X = [A_1 + A_2 ~|~ A_3 + A_4 + A_5] , \\
&\HH_Z = [A_3^T + A_4^T + A_5^T ~|~ A_1^T + A_2^T ] \cdot
\end{align*}
Rows of $\HH_X$ and $\HH_Z$ are respectively the indicator vectors of the $X$ and $Z$ stabilizer generators of the codes~\cite{calderbank1996good, steane1996multiple}.
Each $A_i$ is an $(m\ell)\times (m\ell)$ permutation matrix of the form $Q_{\ell}^u\otimes Q_m^{v}$ with $u \in \{0,1,\dots,\ell-1\}$ and $v \in \{0,1,\dots,m-1\}$. 
Each BB5 code is specified by the values of $\ell, m$ and the matrices $A_1,\dots,A_5$.
\cref{tab:def_BB5} presents two instances of BB5 codes, found by exhaustive search, achieving a larger minimum distance than any BB6 codes with the same parameter $[[n, k]]$.
The construction of $\HH_X$ and $\HH_Z$ from the matrices $A_i$ is identical to Prop.~1 in \cite{voss2024multivariate}, however we allow for matrices $A_i = Q_{\ell}^u\otimes Q_m^v$ whereas the previous construction is restricted to $u=v$.
Note the construction of our $[[48,4,7]]$ code uses $u \neq v$ (see~\cref{tab:def_BB5}).
The paper \cite{voss2024multivariate} introduced a $[[30,4,5]]$ code with stabilizer weight 5 that shares the same code parameters as our $[[30,4,5]]$ BB5 code.
However, \cite{voss2024multivariate} did not report any code with parameters matching our $[[48,4,7]]$ BB5 code.

\begin{table*}
    \centering
    \begin{tabular}{|c|c|c|c|c|c|c|c|}
    \hline
     $[[n,k,d]]$ & $\ell$ & $m$ & $A_1$ & $A_2$ & $A_3$ & $A_4$ & $A_5$   \\
     \hline
     $[[30,4,5]]$ & 5 & 3 & $I_{15}$ & $Q_5\otimes I_3$ & $I_{15}$ & $I_5\otimes Q_3$ & $Q_5^2\otimes Q_3^2$ \\
     \hline
     $[[48,4,7]]$ & 8 & 3 & $I_{24}$ & $Q_8\otimes I_3$ & $I_{24}$ & $I_8\otimes Q_3$ & $Q_8^3\otimes Q_3^2$ \\
     \hline
    \end{tabular}
    \caption{Examples of BB5 codes.}
    \label{tab:def_BB5}
\end{table*}

\begin{figure}
\includegraphics[width=\linewidth]{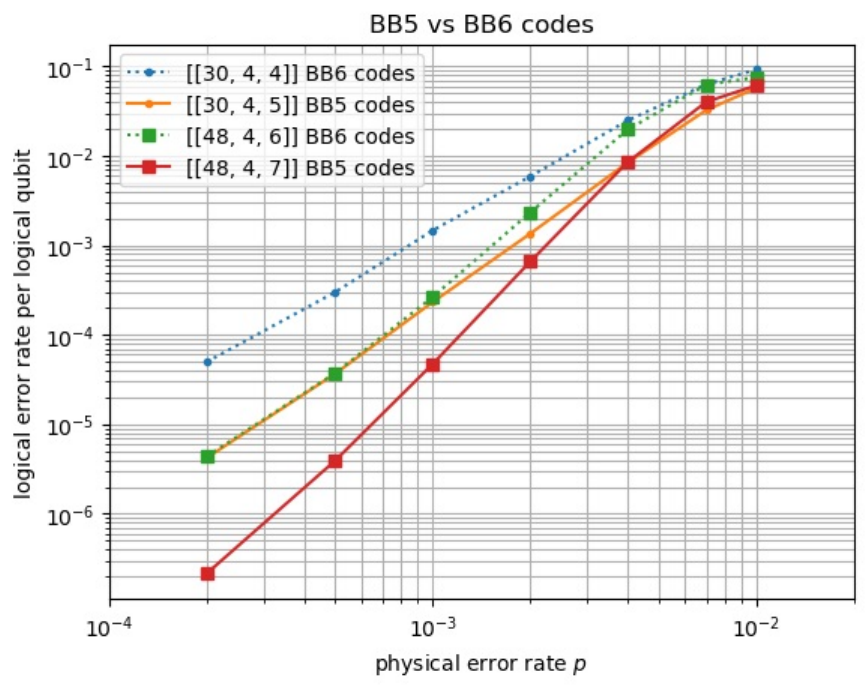}
    \caption{Comparison between BB5 codes and BB6 codes with the same parameters.}
    \label{fig:BB5_vs_BB6}
\end{figure}

To compare BB5 codes against our baseline codes, we estimate their {\em logical error rate}.
In this paper, when we talk about the logical error rate, we always mean the logical error rate per syndrome extraction round and per logical qubit. It is defined as $\frac{q_X + q_Z}{kd}$ where $k$ is the number of logical qubits and $d$ is minimum distance of the code.
The value $q_Z$ is defined to be the probability that a $Z$ logical measurement returns a non-trivial outcome in the circuit that initializes all data qubits in the $\ket{0}$ state, performs $d$ rounds of syndrome extraction, measures all the data qubits and extracts the logical outcomes using a decoder.
The term $q_X$ is defined analogously for $X$ logical measurements, by inserting $H$ gates on the data qubits after initialization and before the final measurement.
We use the ion chain syndrome extraction circuit of \cref{algorithm:ion_chain_syndrome_extraction}. 
The number of ancilla qubits $n_a$ is optimized by \cref{algorithm:syndrome_extraction_tunning_protocol} and reported in \cref{tab:baseline_code_parameters}.

The logical error rate is estimated with a Monte Carlo simulation using Stim~\cite{gidney2021stim}, the BP-OSD decoder~\cite{panteleev2021degenerate, roffe_decoding_2020, Roffe_LDPC_Python_tools_2022} for BB5 and BB6 codes, and PyMatching 2~\cite{Higgott2025sparseblossom} for surface codes.
We use BP-OSD with parameters
\texttt{
max\_bp\_iters = 10\_000, bp\_method = "min\_sum", osd\_order = 5, osd\_method = "osd\_cs"
}.
For all the simulations in this paper, the relative error bar is smaller than $6\%$. 

\cref{fig:BB5_vs_baseline} shows that the $[[48,4,7]]$ BB5 code outperforms all the BB6 codes of our baseline.
While the $[[49,1,7]]$ surface code exhibits a lower logical error rate per logical qubit at low physical error rates, it encodes $4$ times fewer logical qubits than the $[[48,4,7]]$ BB5 code.
Including ancilla qubits, the $[[49,1,7]]$ surface code uses a total of $57$ qubits and the $[[48,4,7]]$ BB5 code uses a total of $54$ qubits.
\cref{fig:BB5_vs_BB6} compares BB5 and BB6 codes with identical $[[n, k]]$ parameters. For both the $[[30,4]]$ and $[[48,4]]$ cases, BB5 codes achieve roughly $6\times$ smaller logical error rate than BB6 codes at physical error rate $10^{-3}$. The gap becomes even larger for smaller physical error rates.

\section{Conclusion}

This paper demonstrates that quantum LDPC codes are well suited for long chain of trapped ions and introduces new quantum LDPC codes that outperform state-of-the-art quantum error correction codes.

Moreover, the high connectivity of ion chains allows us to use fewer ancilla qubits than a superconducting implementation, reducing the qubit overhead, at the price of an increased syndrome extraction time due to less parallelism.

The codes, circuits and optimizations proposed in this work may find applications to other platforms beyond long chains of trapped ions, although our ion chain model does not capture some of the features of these platforms, which may affect the performance of the quantum error correction schemes.

In future work, it would be interesting to explore the performance of other classes of small quantum LPDC codes~ \cite{scruby2024high, lin2024quantum, wang2024coprime, eberhardt2024pruning, lin2025single, guemard2025moderate} under the ion chain model.

For simplicity, we assume that all the single-qubit operations share the same noise rate in the ion chain model. However, in practice, mid-circuit measurements might be noisier than unitary gates~\cite{pogorelov2025experimental}. Noise also varies over qubits and time. In future work, it would be interesting to explore non-uniform noise models including such features.
We can start with a relatively simple observation. If we assign a higher error rate for measurements in our noise model, the optimal number of ancilla qubits $n_a$ will likely increase. This is because the number of measurements is given by $\lceil r/n_a \rceil$ (where $r$ is the number of syndromes) in \cref{algorithm:ion_chain_syndrome_extraction}. Noisier measurements make it beneficial to decrease the total number of measurements, which is achieved by increasing $n_a$.

For simplicity, we assume that all operations take the same amount of time expect measurements. The assumption that measurements are 30 times slower than two-qubit gates is realistic but single-qubit unitary gates are typically much faster~\cite{pogorelov2025experimental}. A future direction is to explore the performance of our quantum error correction scheme when assuming faster single-qubit gates.

It has been observed in \cite{pogorelov2025experimental} that dephasing is the dominant error source on idling qubits in trapped ion hardware, meaning the probability of a $Z$ error is significantly higher than $X$ or $Y$ errors. For the sake of simplicity, our current model uses depolarizing errors for idling qubits, assigning equal probability to $X$, $Y$, and $Z$ errors. We have therefore deferred the incorporation of this physically observed biased noise into our ion chain model for future work. The XZZX variant of surface codes has been shown to perform significantly better than standard surface codes under such biased models \cite{bonilla2021xzzx}, making it an interesting open problem to modify BB5 and/or BB6 codes to similarly enhance their performance under biased noise.

Hadamard gates and controlled-Pauli gates used in \cref{algorithm:ion_chain_syndrome_extraction} are typically non-native gates on a trapped-ion processor.
The decomposition of non-native gates (such as controlled-Pauli and Hadamard gates) into native Mølmer–Sørensen and single-qubit operations, along with subsequent hardware-level optimizations (including gate elimination and commutation), is a substantial field of study in itself \cite{sorensen1999quantum, sorensen2000entanglement, kreppel2023quantum}. To maintain the paper's primary focus on the ion chain model and the new code construction, we have deferred this complex circuit optimization step to future work. Detailed analysis in this area will be necessary for achieving the lowest possible error rates during hardware implementation.

\section{Acknowledgments}

We thank John Gamble, Edwin Tham, Neal Pisenti, Laird Egan, Ricardo Viteri, Alex Ratcliffe, Matthew Boguslawski and Dean Kassmann for insightful discussions during the preparation of this manuscript.

\bibliographystyle{plainnat}
\bibliography{references}

\appendix

\section{Review of surface codes and BB codes}
\label{appendix:surface_code_and_bb_codes}

The section reviews the definition of the surface code and BB codes.

\begin{figure}[h]
    \centering
    \includegraphics[width=0.9\linewidth]{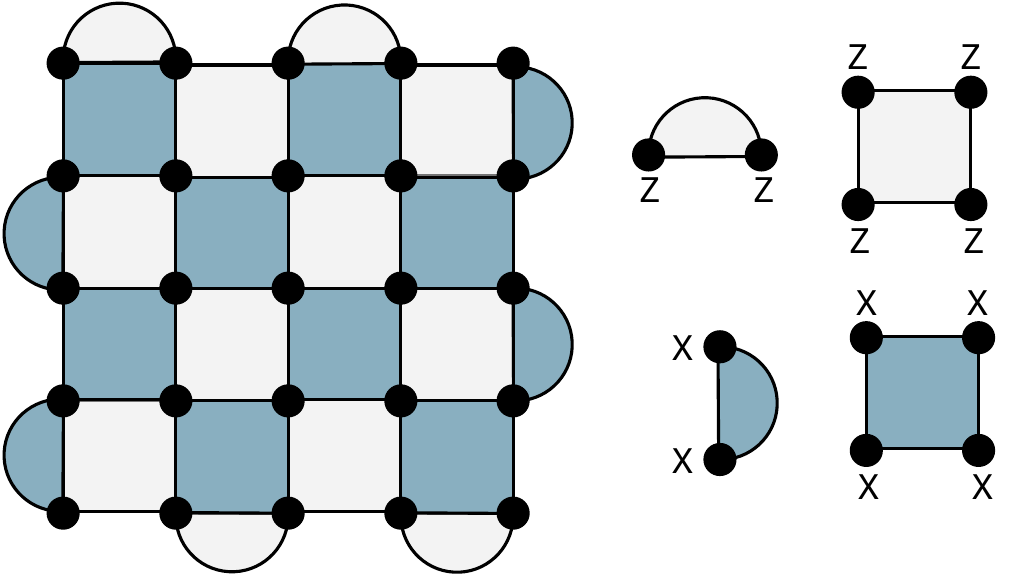}
    \caption{A distance-5 surface code and its stabilizer generators.}
    \label{fig:surface_code_def}
\end{figure}

The distance-$d$ {\em surface code}\footnote{The version of surface code presented in \cref{appendix:surface_code_and_bb_codes} is usually referred to as the rotated surface code in the literature. We use the name surface code throughout the paper for simplicity.} is a $[[d^2, 1, d]]$ stabilizer code defined on a $d \times d$ two-colored square tiling as shown in \cref{fig:surface_code_def}.
Qubits are placed on the $d^2$ vertices and each tile defines a stabilizer generator acting on the four incident qubits either as $X$ or $Z$ depending on the tile color. 
The stabilizer generators associated with boundary tiles act on two qubits only.

Recall from \cref{sec:ion_chain_codes} that $Q_\ell$ is defined as the $\ell \times \ell$ circulant matrix whose first row is $(0 1 0 \dots 0)$.
To introduce BB6 codes, we define the matrices
$
x=Q_{\ell} \otimes I_m
$
and 
$
y = I_{\ell} \otimes Q_m
$
where $m$ and $\ell$ are two positive integers.
A BB6 code is defined by the pair of matrices
\begin{align*}
&\HH_X = [A_1 + A_2 + A_3 ~|~ A_4 + A_5 + A_6] , \\
&\HH_Z = [A_4^T + A_5^T + A_6^T ~|~ A_1^T + A_2^T + A_3^T] ,
\end{align*}
where every $A_i$ is a power of either $x$ or $y$. 
Both $\HH_X$ and $\HH_Z$ have size $(n/2) \times n$. 
The columns of $\HH_X$ and $\HH_Z$ correspond to the $n=2m\ell$ data qubits of the code.
Each row of $\HH_X$ (respectively $\HH_Z$) is the indicator vector of a $X$ (respectively $Z$) stabilizer of the code.
By specifying $m, \ell$ and the matrices $A_1,A_2,\dots,A_6$, we obtain a BB6 code instance.
Given $\ell$ and $m$, there are $(\ell + m)^6$ possible BB6 codes.

\section{Validation of the syndrome extraction tuning protocol}
\label{sec:validate_tuning}

To validate the effectiveness of the syndrome extraction tuning protocol described in \cref{algorithm:syndrome_extraction_tunning_protocol}, we execute this protocol to optimize the number of ancilla qubits $n_a$ of two BB6 codes with parameters $[[30,4,4]]$ and $[[48,4,6]]$ and we check that it provides a sensible choice.

\begin{figure}
\includegraphics[width=\linewidth]{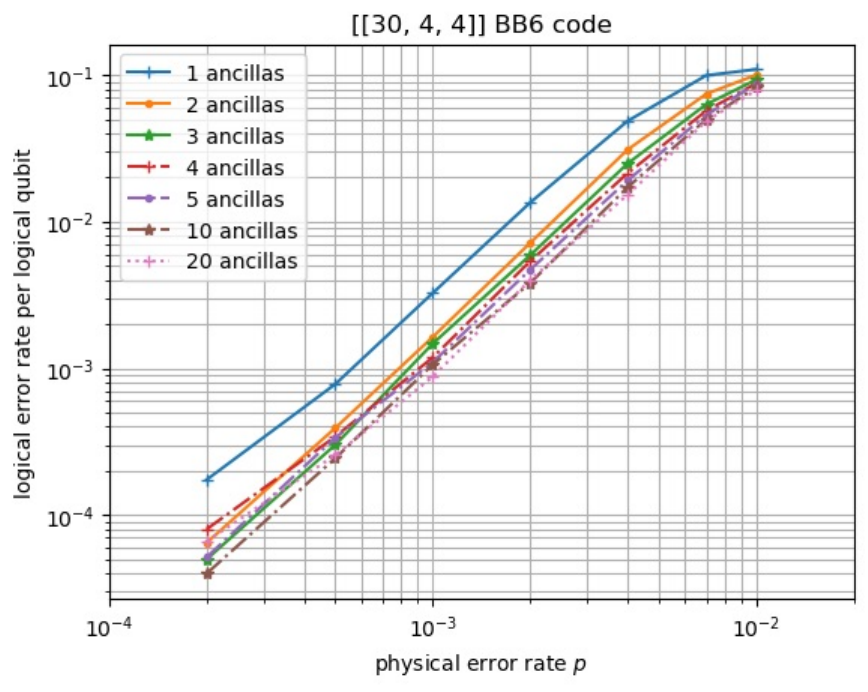}

\includegraphics[width=\linewidth]{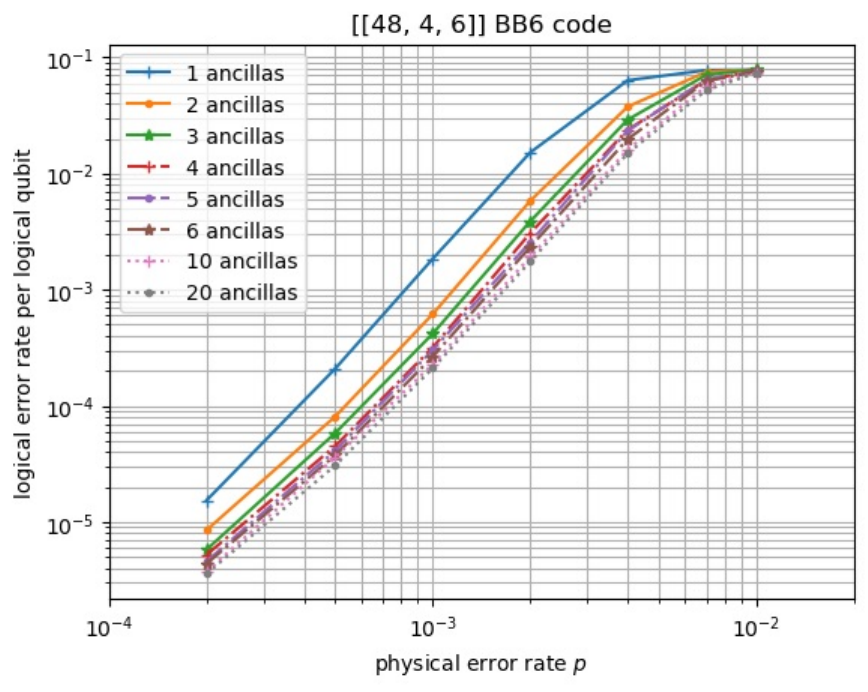}
\caption{\cref{algorithm:syndrome_extraction_tunning_protocol} chooses $n_a=3$ ancillas for $[[30,4,4]]$ BB6 code and $n_a=6$ for $[[48,4,6]]$ BB6 code.}
\label{fig:compare_different_na}
\end{figure}

To confirm that, we estimate the logical error rate of the two codes across a range of ancilla numbers $n_a=1,2,3,4,5,10,20$ as shown in \cref{fig:compare_different_na}.
\cref{algorithm:syndrome_extraction_tunning_protocol} selects $n_a = 3$ for the $[[30,4,4]]$ BB6 code and $n_a=6$ for the $[[48,4,6]]$ BB6 code. 
\cref{fig:compare_different_na} clearly illustrates that increasing $n_a$ beyond these chosen values provides progressively smaller reductions in logical error rates, indicating diminishing returns. 

\section{Fitting formulas for surface codes and BB5 codes}
\label{sec:fitting_formulas}

Most of the code simulations in previous papers employ a circuit model with parallel gate operations and uniform noise rates \cite{fowler2012surface,tremblay2022constant, bravyi2024high, berthusen2025toward}. We will refer to such a circuit model as the parallel circuit model.
Fitting formulas for logical error rates of important code families such as surface codes and BB6 codes were studied under the parallel circuit model \cite{fowler2012surface,bravyi2013simulation, bravyi2024high}. 
For surface codes, \cite{fowler2012surface} presented a remarkably simple fitting formula
\begin{equation} \label{eq:fit_surface}
p_L=c(p/p_{\text{th}})^{(d+1)/2} ,
\end{equation}
where $d$ is the distance of the code, $p_L$ is the logical error rate, $p$ is the physical error rate, $c$ and $p_{\text{th}}$ are constants. For the parallel circuit model,  the constants are $c=0.03$ and $p_{\text{th}}=0.0057$ \cite{fowler2012surface}. Note that the same formula \eqref{eq:fit_surface} with the same constants $c$ and $p_{\text{th}}$ effectively approximates the logical error rate of all surface codes. In contrast, \cite{bravyi2024high} uses a more complicated formula $p_L=p^{d/2}e^{c_0+c_1p+c_2p^2}$ to fit BB6 codes, and each BB6 code instance requires a distinct set of constants $c_0,c_1,c_2$.

Under the ion chain model, we find that the formula \eqref{eq:fit_surface} with constants $c=0.003$ and $p_{\text{th}}=0.0032$ provides a good fit for the 3 surface codes in \cref{tab:baseline_code_parameters}. However, this formula does not generalize to longer surface codes with distance larger than 7. For BB5 codes, we use a more complicated fitting formula $p_L=p^{(d+1)/2}e^{c_0+c_1p+c_2p^2}$, where each BB5 code instance has its own set of constants $c_0,c_1,c_2$. These constants for the two BB5 code instances are listed in \cref{tab:fitting_BB5}.

\begin{table}[h]
    \centering
    \begin{tabular}{|c|c|c|c|}
    \hline
     $[[n,k,d]]$ & $c_0$ & $c_1$ & $c_2$   \\
     \hline
     $[[30,4,5]]$ & 12.869  & -340.43  &  15878  \\
     \hline
     $[[48,4,7]]$ & 18.256  & -260.44 & 680.65  \\
     \hline
    \end{tabular}
    \caption{Constants in the fitting formula $p_L=p^{(d+1)/2}e^{c_0+c_1p+c_2p^2}$ for BB5 codes.}
    \label{tab:fitting_BB5}
\end{table}

\end{document}